\documentclass[10pt,aps,twocolumn,pra,nofootinbib,superscriptaddress,floatfix,reprint,preprintnumbers,frontmatterverbose] {revtex4-1}

\usepackage{amssymb,amsmath}
\usepackage{subfig}
\usepackage{graphicx}
\usepackage{color}
\usepackage{hyperref}
\usepackage{listings}
\usepackage{csquotes}

\graphicspath{{figures/}}

\begin{document}

\title{Effect of higher-order interactions in a hybrid Electro-Optomechanical System}

\author{Vinay Shankar} 
\affiliation{Department of Physics, Shiv Nadar University, Greater Noida, Uttar Pradesh 201314, India}
\author{Suneel Singh}
\affiliation{School of Physics, University of Hyderabad, Hyderabad, Telangana 500046, India} 
\author{P. Anantha Lakshmi}
\email{palsp@uohyd.ernet.in}
\affiliation{School of Physics, University of Hyderabad, Hyderabad, Telangana 500046, India}
 
\date{\today}

\begin{abstract}
A detailed analysis highlighting the effect of optomechanical non-linearity on the dynamical evolution of a hybrid electro-optomechanical system (EOMS) is presented. The study, conducted over a wide range of parameter regime reveals that the quadratic coupling term significantly alters the dynamics of the system and thus cannot be ignored for any potential applications.   

\end{abstract}
\maketitle
\section{\label{sec:one}Introduction}
The interest in cavity optomechanics (OM) has seen a big spike in recent years. Experimental realizations of such systems \cite{aspelmeyer2014cavity} have considered a wide range of masses for mechanical elements, ranging from a few nanograms \cite{groblacher2009observation} all the way to a few kilograms \cite{cuthbertson1996parametric}. Many features have been observed viz., the normal-mode splitting \cite{dobrindt2008parametric,huang2009normal,huang2010normal}, optomechanically induced transparency (OMIT)  \cite{weis2010optomechanically,agarwal2010electromagnetically,safavi2011electromagnetically,xiong2012higher,sohail2016optomechanically}, sideband cooling \cite{mancini1998optomechanical,wilson2007theory,marquardt2007quantum,schliesser2008resolved,teufel2011sideband,chan2011laser,chen2015cooling,zeng2017ground}, and mechanical squeezing \cite{safavi2013squeezed,purdy2013strong,wang2016steady}, across different parameter regimes. To achieve greater control over these features for potential applications, additional hybrid elements are added to the OM system. Hybrid elements like multi-level atoms \cite{genes2011atom,breyer2012light,yi2014ground,kampschulte2014electromagnetically,pirkkalainen2015cavity,jiang2017fano,he2018multiple,liu2018generation,kong2018two}, Bose-Einstein condensate (BEC) \cite{zhang2010hamiltonian}, charged objects \cite{zhang2012precision,xiong2017precision,xiong2017highly,kong2017coulomb}, quantum well \cite{yellapragada2018optomechanical,wang2017tunable}, to name a few, have been incorporated and widely studied. Many earlier studies have also used coherent mechanical pumps to boost the OM coupling in hybrid OM systems, thus enhancing low-coupling regime features \cite{ma2015optomechanically,xu2015controllable,sun2017optical,suzuki2015nonlinear,jiang2016phase,yellapragada2018optomechanical}.      
\par
In the present study, a hybrid Electro-optomechanical system (EOMS) is considered. In earlier works on EOMS, features analogous to those observed in traditional OM systems like microwave-controlled OMIT, optomechanically induced absorption (OMIA) \cite{wu2018microwave,qu2013phonon} and higher-order sideband generation \cite{chen2016second,si2018tunable} have been reported. One of the major applications for such a system is that of a bi-directional microwave to optical convertor. These type of convertors have significant potential for application in the fields of quantum information and quantum communications. This has been made feasible since the mechanical mode acts as a link between the optical and microwave modes, which are otherwise uncoupled. Recent research on such convertors is addressed in references \cite{andrews2014bidirectional,tian2015optoelectromechanical,vainsencher2016bi,xu2016nonreciprocal,tian2017nonreciprocal,forsch2019microwave}. 
\par
Recent theoretical and experimental studies have highlighted the promising effects of quadratic coupling in OM systems. The \enquote{membrane in the middle} configuration \cite{jayich2008dispersive} or usage of ultracold atoms \cite{purdy2010tunable} were some of the ways to achieve a quadratically coupled OM system. Features like OMIT \cite{huang2011electromagnetically}, mechanical squeezing \cite{nunnenkamp2010cooling}, photon blockade \cite{liao2013photon}, ground-state cooling \cite{yang2019ground}, optical amplification \cite{wang2019tunable}, slow light \cite{zhan2013tunable}, optomechanically induced opacity \cite{si2017optomechanically} and also a recent proposal for a highly sensitive mass sensor \cite{liu2019realization} have been reported for these systems. Additionally, Karuza et al.\cite{karuza2012tunable} and Xuereb et al.\cite{xuereb2013selectable} have demonstrated methods of tuning the sign of the quadratic coupling through tilting of the membrane at an angle and positioning the dielectric spheres at the node or antinode of the cavity, respectively. While the above mentioned features arose by considering purely quadratic OM interaction, similar features like OMIT \cite{zhang2018optomechanically}, optical bistability \cite{nejad2017effect}, optical and mechanical squeezing \cite{kumar2015effects,satya2017squeezing}, and normal mode splitting at lower pump powers \cite{satya2019mimicking}, have been reported in systems considering both linear and quadratic coupling terms. Dalafi et al.\cite{dalafi2018effects} looked at the effects of adding the quadratic term in an OM system coupled with BEC.     
\par
Chaos and non-linear behaviour have been studied in a wide variety of areas with numerous applications, in particular, in the functioning of random number generators \cite{gleeson2002truly} and in encrypted communications \cite{sivaprakasam1999signal,sivaprakasam2000message,sciamanna2015physics}. In OM systems, there have been numerous reports detailing the emergence of chaos in both standard and hybrid systems through theoretical and experimental means \cite{carmon2007chaotic,zhang2010hamiltonian,larson2011photonic,ma2014formation,lu2015p,bakemeier2015route,wurl2016symmetry,navarro2017nonlinear}. The key objective to achieve, while studying such systems, is to be able to manipulate/control the random behaviour for potential applications and some of the above cited studies have attempted to address this question.Occurrence of chaos in typical OM systems necessitates the use of large pump powers which is not feasible for such systems consisting of nanomechanical mirrors. Adding a hybrid element would provide a means to control the system while also keeping the system parameters within experimental bounds.
\par
Recently, Wang et al.\cite{wang2016controllable} reported chaos in a hybrid EOMS, wherein they have considered OM interactions up to first order. One of the key features of their proposal was that the microwave mode and optical mode are indirectly coupled through the mechanical resonator, thereby keeping the control and generation points distinct. 
In this context it would be of interest to examine the non-linear effects in such systems as generally, chaos is known to be highly sensitive to a system's initial conditions. 
With this motivation, we carry out a study of non-linear dynamics in a hybrid EOMS with the inclusion of an additional quadratic interaction term. A detailed analysis of how the non-linearity affects the interplay between the microwave and optical fields and the system dynamics is presented.
\par
The paper is structured as follows. The system configuration along with its dynamics, i.e., Hamiltonian and the equations of motion are presented in Section~\ref{sec:two}. In order to distinguish between the effects arising due to the microwave field and the optical fields (with the inclusion of the quadratic interaction term), the study is presented in two different sections. Section~\ref{sec:three} discusses the results of the microwave field effects, reporting the onset of chaos as well as the sensitivity of the system's response to the microwave field, while section~\ref{sec:four} explores the effect of the optical fields on the system dynamics. Summary and conclusions are presented in  section~\ref{sec:five}.

\section{\label{sec:two}System Framework}
\begin{figure}[h!]
\centering
\includegraphics[width=1.0\linewidth]{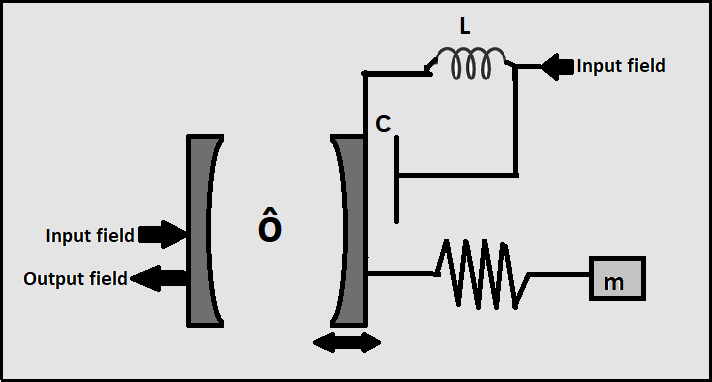}
\caption{Schematic of a hybrid Electro-Optomechanical system (EOMS)}
\label{fig:System}
\end{figure}
A hybrid EOMS, a schematic of which is shown in Figure~\ref{fig:System}, consisting of a mechanical mode (oscillator) coupled to an optical mode (Fabry-Perot cavity) and a microwave mode (LC circuit), is considered for the present study. Control fields from both ends are used to drive the cavity and the LC circuit respectively. The Hamiltonian, incorporating the various interactions present in the system, can be written as $H^{tot} = H^0+ H^{1}$ where the first term accounts for the energies of each of the sub-systems and the second term describes the interactions between different sub-systems considered here.
\begin{align}
    \begin{split}
       &H^{0}= \hbar \omega_o\,o^{\dagger} o +\hbar \omega_a\,a^{\dagger} a +\frac{p^2}{2m}+\frac{1}{2} m w_m^2q^2\\
       &+i\hbar \epsilon_o(o^{\dagger}e^{-\omega_{ol} t}-oe^{i\omega_{ol}t}) +i\hbar \epsilon_a(a^{\dagger}e^{-\omega_{al} t}-ae^{i\omega_{al}t})\\
       &H^{1}= H^{1}_{O-M} +H^{1}_{MW-M}\\
       &{\text where}\\
       &H^{1}_{O-M}=- \hbar \alpha_{lin} \,o^{\dagger}o q+\hbar \alpha_{quad} \,o^{\dagger}o q^2,\\
       &H^{1}_{MW-M}= -\hbar \beta \,a^{\dagger}a q  \\ 
    \end{split}
\label{eq:Hamiltonian}
\end{align}

The first two terms in $H^0$ refer to the cavity and microwave mode energies, with frequencies $\omega_o$ (fundamental mode) and $\omega_a =\frac{1}{\sqrt{LC}}$ respectively. The $L$ and $C$ appearing in the above expression are the inductance and capacitance of the microwave circuit. $o$ ($a$) and $o^{\dagger}$ ($a^{\dagger}$) are the annihilation and creation operators of the cavity (microwave) field, respectively. The third and fourth terms in $H^0$ together represent the energy of the mechanical mode (mirror), where $q$ and $p$ represent the position and momentum coordinates of the mirror. The fifth (sixth) term represents the field driving the cavity (microwave circuit) with frequency $\omega_{ol}$($\omega_{al}$) and amplitude $\epsilon_o$($\epsilon_a$). Here $\epsilon_0=\sqrt{\frac{2 \kappa_o P_o}{\hbar \omega_{ol}}}$ where $P_o$ is the power of the laser and $\kappa_o$ is the cavity decay rate. Similarly, the driving amplitude for the microwave field $\epsilon_a=\sqrt{\frac{2 \kappa_a P_a}{\hbar \omega_{al}}}$ where $P_a$ and  $\kappa_a$ are the corresponding power and decay rates, respectively. 
\par
The interaction energy represented by $H^{1}$, consists of two contributions arising due to the OM and the micro-mechanical couplings. The OM coupling $H^{1}_{O-M}$, consists of linear and quadratic contributions with their respective coupling constants $\alpha_{lin}$ and $\alpha_{quad}$. The final term in the interaction Hamiltonian, $H^{1}_{MW-M}$, represents the micro-mechanical coupling, with a coupling constant $\beta$. The Hamiltonian is considered in a frame rotating with the frequencies $\omega_{al}$ (MW mode), $\omega_{ol}$(optical), thus removing the fast oscillations.  
\par
Taking into account damping and decay processes, the Heisenberg-Langevin equations of motion are obtained as 
\begin{align}
    \begin{split}
        &\dot{q}=\frac{p}{m}\\
        &\dot{p}=-\gamma_m p-m\omega_{m}^2 q+\hbar(\alpha_{lin}-2 \alpha_{quad}\,q)(o_{r}^2+o_{i}^2)+\hbar \beta (a_{r}^2+a_{i}^2)\\
        &\dot{o_r}=-\kappa_o o_r+(\Delta_o-\alpha_{lin}\,q+\alpha_{quad}\,q^2)o_i+\epsilon_o\\
        &\dot{o_i}=-\kappa_o o_i-(\Delta_o-\alpha_{lin}\,q+\alpha_{quad}\,q^2)o_r\\
        &\dot{a_r}=-\kappa_a a_r+(\Delta_a-\beta\,q)a_i+\epsilon_a cos(\phi)\\
        &\dot{a_i}=-\kappa_a a_i-(\Delta_a-\beta\,q)a_r-\epsilon_a sin(\phi).\\
    \end{split}
\label{eq:Langevin Equations}
\end{align}
In the above, $o_r\,(a_r)$ and $o_i\,(a_i)$ are the real and imaginary parts of the optical (microwave) field respectively, such that, $<{\hat{o}}>=\,o_r+i\, o_i$ and $<{\hat{a}}>=\,a_r+i\, a_i$. Here $\Delta_o=\omega_o-\omega_{ol}$ ($\Delta_a=\omega_a-\omega_{al}$) is the optical(microwave) detuning and $\phi$ is the relative phase between the optical and microwave driving fields. The quantity of interest here is the optical intensity $I_o$ given by $I_o=o_{r}^2+o_{i}^2$, which can be obtained by solving the system of differential equations for $o_r$ and $o_i$. It would also be of interest to study the dynamical evolution of $I_o$, derived using the above equations as
\begin{equation}
    \frac{dI_o}{dt}=-2\kappa_o I_o+2\epsilon_o o_r.
\end{equation}

We have obtained a set of coupled non-linear differential equations (Eq.~\ref{eq:Langevin Equations}) that describe the time evolution of the system. Determining the analytical solution of such a system of equations is highly cumbersome. Instead, numerical simulations of this set of differential equations, over a sufficient time interval, for an exhaustive parameter range will be carried out, which will  provide valuable insights into the dynamical behaviour of the system.  
\par
Using the \emph{ode45} function, which utilises the fourth-order Runge-Kutta (R-K) method, the optical intensity ($I_o$) as a function of time and the time rate of change of intensity $\dot{I}$ is calculated, giving the phase space plot ($\dot{I} \, vs\, I$). The initial condition for each of the system variables are set to zero and time evolution of the system is studied, results of which are presented in the next section. 

\section{\label{sec:three}Microwave field variation}
In this section, a detailed study of the system dynamics is presented at a relatively low optical power ($P_o =0.5 mW$), over a range of microwave powers, \textit {with inclusion of the quadratic OM interaction term}.  
\subsection{Emergence of Chaos and Sensitivity}
\begin{figure}[h!]
\centering
\includegraphics[width=1.0\linewidth]{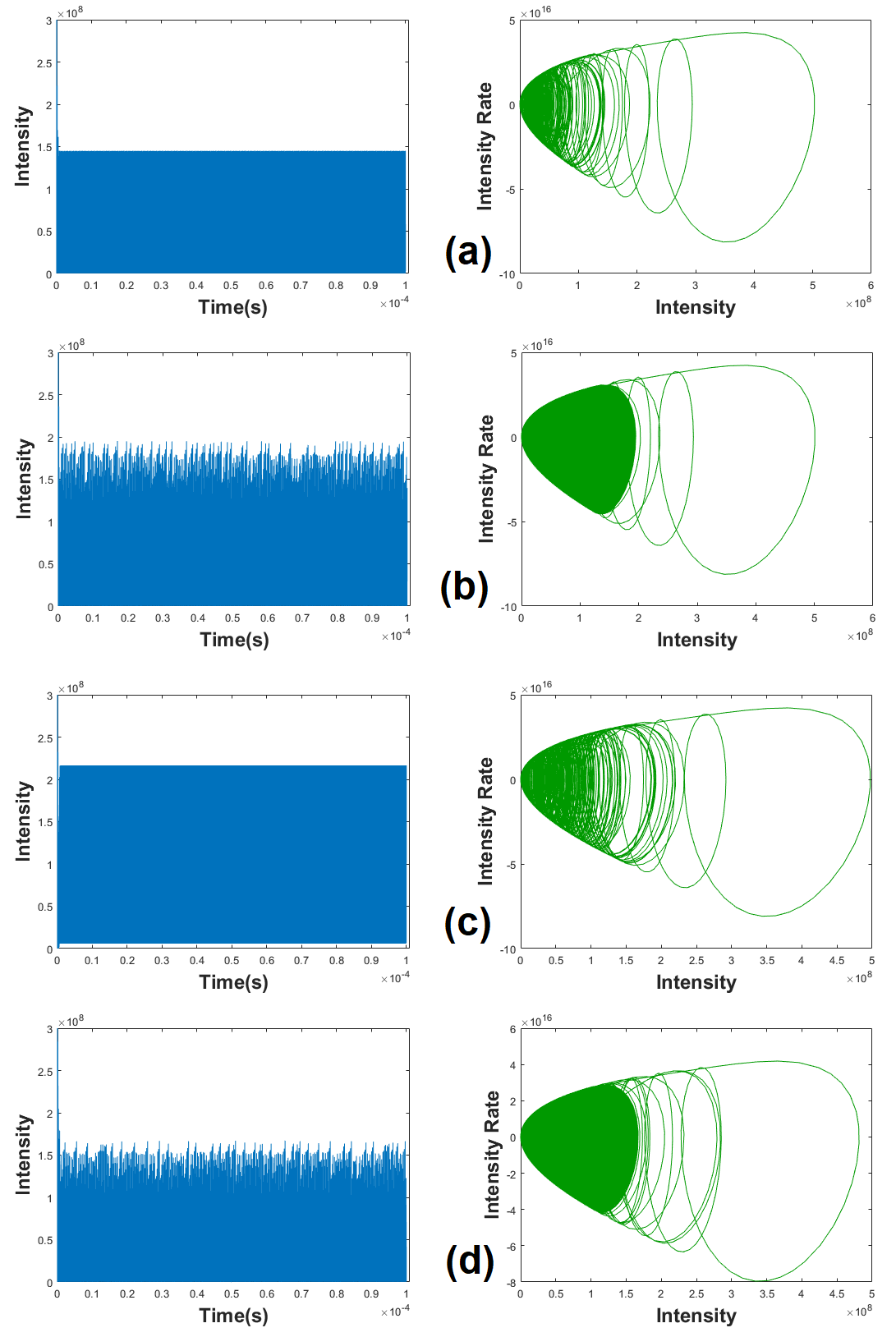}
\caption{The intracavity field intensity $I_o \,vs\, time$ (left-blue) and the optical phase space trajectory $\dot{I}\,vs\, I$ (right-green), over a time interval of $[0, 100\mu s]$, for MW driving powers (a) $P_a =5.90 \,\mu W$, (b) $P_a =5.97\,\mu W$, (c) $P_a =6.09\,\mu W$, (d) $P_a =6.67\,\mu W$. 
The other parameters are fixed at $m = 105\,ng$, $P_o =0.5 \,mW$, $\Delta_o =\omega_m$, $\Delta_a =0$, $\omega_m =73.5 \,MHz$, $\gamma_m =1\,MHz$, $\omega_o = 1.9\,GHz$, $\kappa_o =0.4 \,\omega_m$, $\omega_a =1\,THz$, $\kappa_c = 0.8 \,\omega_m$, $\alpha_{lin} = 5.6 \times 10^{18}\,Hz/m$, $\frac{\alpha_{quad}}{\alpha_{lin}} = 10^{-6}$ and $\beta =\alpha_{lin}$}
\label{fig:Emergence of Chaos}
\end{figure}

To start, the relative coupling strength defined as the ratio of the the quadratic coupling constant to the linear coupling term is fixed at $\frac{\alpha_{quad}}{\alpha_{lin}} = 10^{-6}$. For low driving powers, the system exhibits periodic behaviour as demonstrated in Figure~\ref{fig:Emergence of Chaos}-(a). At $P_a =5.97\,\mu W$, the first occurrence of random behaviour is seen in the system (Figure~\ref{fig:Emergence of Chaos}-(b)). For a small incremental change in power to $P_a =6.09 \,\mu W$, the system tends to periodic behaviour again (Figure~\ref{fig:Emergence of Chaos}-(c)) before displaying chaotic behaviour at $P_a=6.67\,\mu W$ ((Figure~\ref{fig:Emergence of Chaos}-(d)) . These results are found to be in agreement with that of the linear case ($\alpha_{quad}=0$) as expected, as the non-linear effects will start to emerge only for higher powers of the microwave driving field. Alternatively, the power of the optical field can be boosted to observe any non-linear effects and this idea is elaborated in Section~\ref{sec:four}. 
\par
\begin{figure}[h!]
\centering
\includegraphics[width=1.0\linewidth]{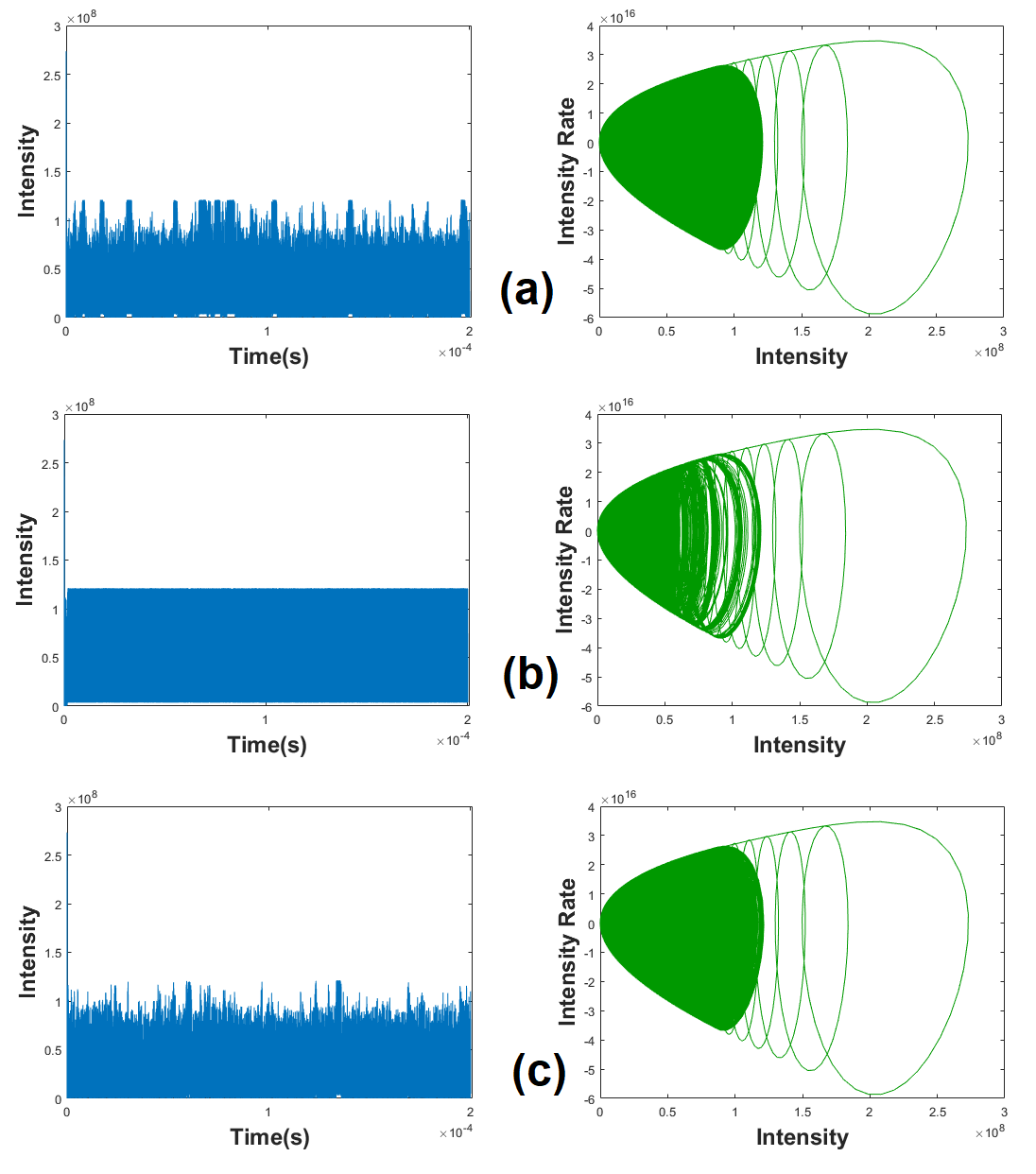}
\caption{The intracavity field intensity $I_o \,vs\, t$ (left-blue) and the optical phase space trajectory $\dot{I}\,vs\, I$ (right-green), over a time interval of $[0, 200\,\mu s]$, for Microwave driving powers (a) $P_a=30.90\,\mu W$, (b) $P_a=30.91\,\mu W$, (c) $P_a=30.98\,\mu W$, (d) $P_a=31.07 \,\mu W$ for a time interval of $[0, 200\,\mu s]$. The other parameters are the same as in Fig~\ref{fig:Emergence of Chaos}}
\label{fig:Sensitivty case}
\end{figure}

This study has been repeated over a wide range of microwave powers and it is observed that the system goes through multiple cycles of chaos - order - chaos throughout, as demonstrated in Figure~\ref{fig:Sensitivty case}. Incremental changes in microwave field required in order to see such fluctuations vary from as low as $0.01 \,\mu W$ in some cases, to a few $\mu W$. The system is thus highly responsive to the microwave power, which complements the fact that the microwave field has a significant impact on the nature of the OM coupling. The precise form of such an influence cannot be ascertained as it is not possible to obtain an analytical solution of the system of equations. 

\subsection{Change in chaotic lifetime}
The period for which a system exhibits random behaviour is termed as its \textit{chaotic lifetime} and such a behaviour is referred to as \textit{transient chaos}\cite{tel2015joy}. For certain microwave powers, the system is seen to exhibit a finite lifetime of chaos, before settling into periodic behaviour. Since the focus here is to explore the regime where non-linear effects become significant, the results are presented for each case, with and without the inclusion of the quadratic coupling term.   

\begin{figure}[h!]
\centering
\includegraphics[width=1.0\linewidth]{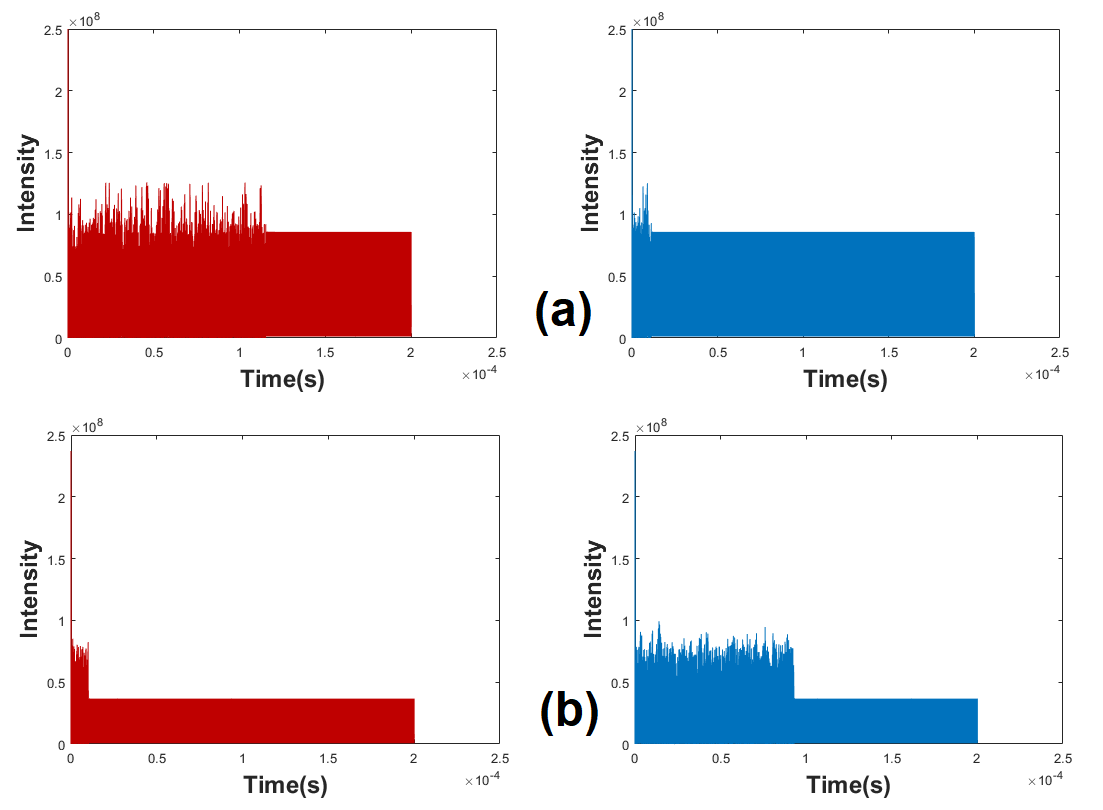}
\caption{The intracavity field intensity $I_o \,vs\, time$ is shown for the linear coupling case (left-red) and non-linear case (right-blue) at different microwave powers- (a) $P_a=28.6\,\mu W$, (b) $P_a=45.1\,\mu W$ over a time interval of $[0, 200\,\mu s]$. The other parameters are the same as in Fig~\ref{fig:Emergence of Chaos}}
\label{fig:Change in lifetime}
\end{figure}

Though numerous cases were observed, where the quadratic term affects the chaotic lifetime, however, in Figure~\ref{fig:Change in lifetime} we show the results only for those parameters, where a significant change is evident. These results show that the quadratic term does not have a uniform influence on the system dynamics. For instance, Figure~\ref{fig:Change in lifetime}-(a) shows a decrease in lifetime at $P_a=28.6\,\mu W$, whereas in Figure~\ref{fig:Change in lifetime}, an increase in chaotic lifetime is observed. This non-uniform behaviour can be attributed to the fact that the inclusion of the second-order coupling term increases the non-linear nature of the interaction, thus altering the system dynamics in a pronounced manner. These results indicate that the quadratic term cannot be ignored even for coupling ratios as small as $10^{-6}$.

\subsection{Effect of relative phase between the fields}

\begin{figure}[h!]
\centering
\includegraphics[width=1.0\linewidth]{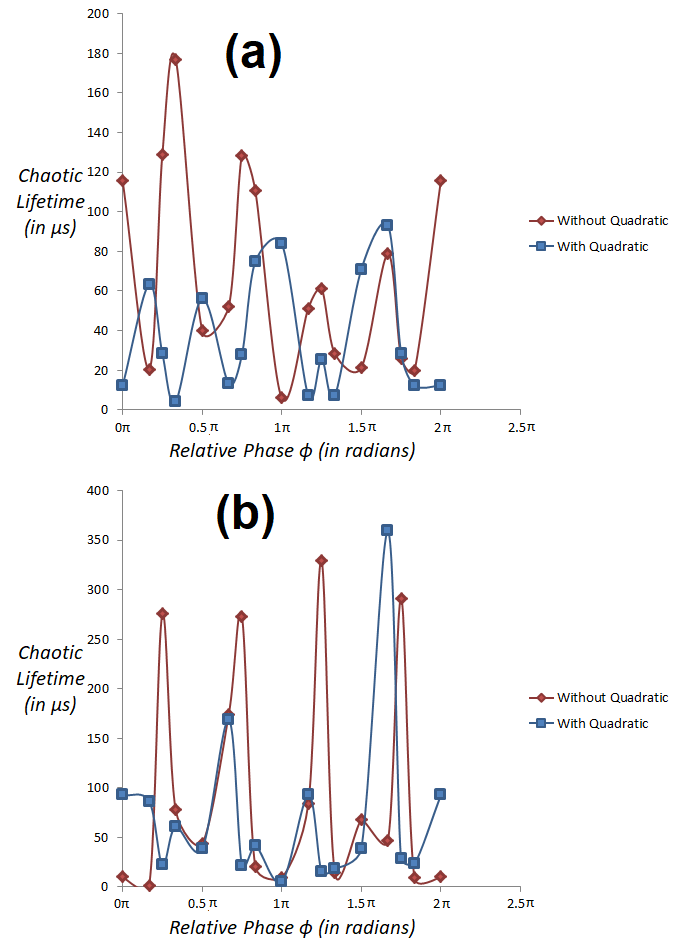}
\caption{The chaotic lifetime (in $\mu s$) is shown as a function of the relative phase between optical and microwave fields ($\phi$) for linear (red line-diamond) and non-linear case (blue line-square) at different microwave powers- (a) $P_a=28.6 \,\mu W$ , (b) $P_a=45.1 \,\mu W$. The other parameters are the same as in Fig~\ref{fig:Emergence of Chaos}}
\label{fig:Effect of Phase on the lifetime}
\end{figure}
Upon transforming the Hamiltonian, the relative phase $\phi$ between the optical and microwave pump lasers enters the equations of motion as shown in Equation~\ref{eq:Langevin Equations}. In the previous section, the relative phase was kept fixed at zero. The focus now, is on the change in the chaotic lifetime as the relative phase between the fields is varied. Figure~\ref{fig:Effect of Phase on the lifetime} shows the variation in lifetime with respect to the relative phase, for both the linear and non-linear cases. There is a significant difference in the lifetime between the two curves, corresponding to the cases where the quadratic term is absent/present, as the phase varies from $[0,2\pi]$. Overall, we observe that the variation in phase has a significant effect on the system dynamics in the presence of the quadratic non-linearity.
\par
Mention needs to be made that the Lyapunov exponent for different cases did not provide conclusive evidence of chaos and this is attributed to the possibility that Perron effects \cite{leonov2007time} are playing a role. This claim is not verifiable due to unavailability of exact solutions for this system. 
\section{\label{sec:four} Optical field variation}

\begin{figure}[h!]
\centering
\includegraphics[width=1.0\linewidth]{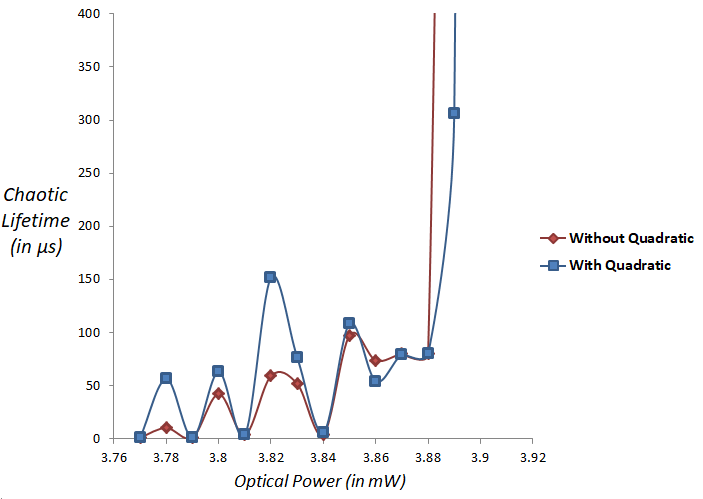}
\caption{The chaotic lifetime (in $\mu s$) is shown as a function of the optical power ($P_o$) for linear (red line-diamond) and non-linear case (blue line-square). The other parameters are the same as in Fig~\ref{fig:Emergence of Chaos}} with $\phi=0$.
\label{fig:Optical power effects}
\end{figure}

In this section, the non-linear effects arising out of the quadratic coupling term is studied, as the optical power $P_o$ is increased, holding the microwave power fixed at $P_a=9\,\mu W$. In the previous section, it was observed that the quadratic term influences the chaotic lifetime for optical powers as low as $P_o=0.5\,mW$.
\par
For $P_a=9\,\mu W$, at relatively small optical powers, the system exhibits transient behaviour for a particular lifetime, before settling into periodic behaviour. Until an optical power of approximately $P_a=3.7\,mW$, the system remains periodic, beyond which it shows a finite chaotic lifetime. As seen in Figure~\ref{fig:Optical power effects}, above certain threshold value of the optical power, the quadratic term has a strong influence on the system dynamics. There is a significant change in the chaotic lifetime, when the quadratic term is absent (present) shown in red (blue) curves. Beyond $P_a=3.88 \,mW$, when the quadratic term is absent, the chaotic lifetime is very large indicating that the system has settled into chaos. For the case with the quadratic term included, the system still shows a finite lifetime, before eventually approaching large lifetimes, when the optical power is increased beyond $3.9\,mW$. This difference shows that, with the quadratic term included, the system requires a higher threshold of optical power to settle into chaos. 

\begin{figure}[h!]
\centering
\includegraphics[width=1.0\linewidth]{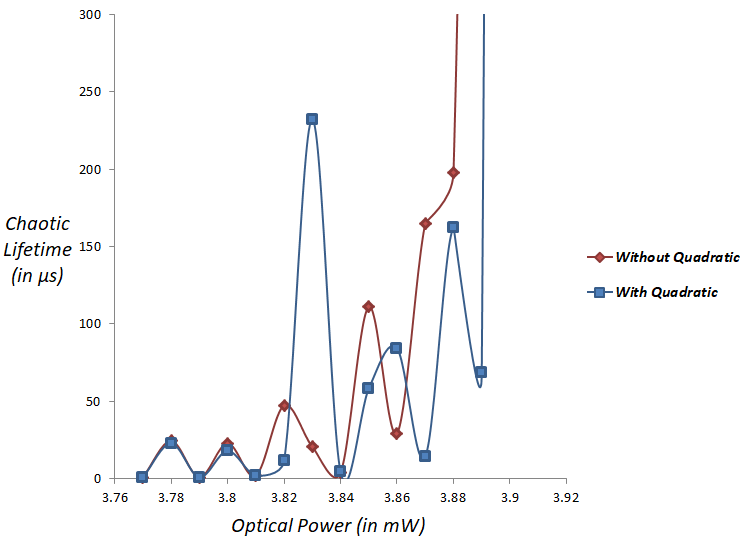}
\caption{The chaotic lifetime (in $\mu s$) is shown as a function of the optical power ($P_o$) for linear (red line-diamond) and non-linear case (blue line-square) for relative phase $\phi=\frac{3\pi}{2}$. The other parameters are the same as in Fig~\ref{fig:Emergence of Chaos}}.
\label{fig:Phase effects with Optical power}
\end{figure}

Another parallel study conducted, is the effect of tuning the relative phase between the optical and microwave fields. Figure~\ref{fig:Phase effects with Optical power} shows for a value of $\phi=\frac{3\pi}{2}$. Once again, the change in chaotic lifetime due to the presence/absence of the quadratic term is significant. 

In the case of $\phi=0$ (Figure~\ref{fig:Optical power effects}), the blue (red) curve corresponding to the quadratic term being present (absent), are in phase, whereas, we observe that in the case of $\phi=\frac{3\pi}{2}$ shown in Figure~\ref{fig:Phase effects with Optical power}), the blue curve leads the red curve. This difference is solely due to the effect of relative phase between the fields on the quadratic coupling term. To reiterate, all these results have been presented for the ratio between the coupling constants ($\frac{\alpha_{quad}}{\alpha_{lin}}$) fixed at $10^{-6}$. 

All these results indicate that, even for a very small coupling ratio, the quadratic term has a significant impact on the system dynamics in different parameter regimes. Therefore one cannot neglect the effect of nonlinearity (quadratic coupling) for any potential applications of the system such as in random number generators and encrypted communications. 

\section{\label{sec:five}Conclusion}
This study explores the parameter regime over which the non-linear effects become prominent, leading to significant changes in chaotic behaviour of a hybrid EOMS. A wider search of the parameter space has revealed that the onset of chaos, occurs at a considerably lower microwave power in the presence of the quadratic coupling, as opposed to the absence of the same. In particular, it is observed that chaotic lifetimes are significantly altered when the non-linear effects are taken into account. The study presented here shows beyond a doubt that, the effect of quadratic coupling can not be ignored when one is studying features which depend sensitively on the system parameters, in this case chaos.  
 
\section{Acknowledgement}
Vinay Shankar would like to thank Dr. Santosh Kumar, Shiv Nadar University for the discussions and valuable inputs during the study.
\bibliography{References.bib}
\end{document}